\documentstyle[12pt]{article}
\title{\bf{Born--Infeld theory of gravitation: Spherically symmetric
static solutions}}
\author{Dmitriy Palatnik\\SMG Marketing Group, 875 N. Michigan Ave.,
Chicago, IL 60611, USA}

\begin{document}

\renewcommand{\theequation} {\arabic{section}.\arabic{equation}}

\maketitle 

\section{Introduction}

In the work \cite{9608014} a theory of electroweak
and gravitational fields based on
the Born--Infeld type of action was suggested. In this paper
the attention is narrowed to the gravitational sector only;
respectively, all fields, except gravitational, and matter are considered
to be absent. In the third section the modified vacuum
Einstein equations are derived  from the Born--Infeld action. 
In the fourth section the equations for the static
spherically symmetric case are considered in a more detail. 
The asymptotics for the Schwarzschild solution as a decomposition
in parameter $L = (20/3)^{1\over2}(k^{1\over2}\hbar)/(ec) 
\approx 10^{-32}$ cm is obtained. 
In the fifth section an interior solution is obtained, 
also with static, spherically symmetric spacial geometry 
corresponding to the region $r \in [0,\,r_0)$ with $r_0 \propto L$.
 
\section{Motivation and basic notions}
\setcounter{equation}{0}

More than sixty years ago Born and Infeld suggested a theory
\cite{BI} of electromagnetism with non-quadratic lagrangian.
The action of fields,
\begin{equation}  \label{original}
S_{BI} \,= \, {\int}d\Omega\left(\sqrt{- \det | \eta_{ab} +
f_{ab} |} - \sqrt{- \det | \eta_{ab} |}\,\right),
\end{equation}
leads to non-linear Maxwell equations for dimensionless tensor of
electromagnetic field, $f_{ab}$. Here tensor $\eta_{ab}$ represents
flat spacetime metric. Authors were inspired by a certain result
of their theory, namely, by absence of the singularity in
a special case of static, spherically symmetric electric field.
One may use the same approach, treating gravitation. As in the case
of electromagnetic fields, one may expect that non-linear (in curvature
tensor) corrections to the Einstein equations should cancel with linear term
on small distances in such manner, that singularity (in the case of
the Schwarzschild solution) disappears. Accepting the idea of non-linear
Einstein equations (in a sense given above),
one has a lot of alternatives for selecting different powers of, say,
the Riemann
tensor (contracted with metrics) with arbitrary coefficients. So, one
appears to be helpless to select the unique lagrangian. 
In this respect Born--Infeld approach
suggests a method for construction of (almost) unique  
lagrangian, by using
the only dimensional parameter, which happens to be a characteristic
length, $L \approx 10^{-32}$ cm.

One can further decrease the number of competing theories, by demanding that
the theory should be
compatible with quantum field theory. This means the following.
The Dirac action for a fermion (described by Dirac's 4--spinor, $\psi$),
doesn't contain the metric tensor, $g_{ab}$, and the Christoffel symbol,
$\Gamma_{ab}^{c}$, as fundamental elements. Instead, the basic notions,
from which geometry can be derived, are the set of Dirac matrices,
$\gamma_a$, satisfying relations, 
\begin{equation}
\gamma_{(a}\,\gamma_{b)} \,=\, g_{ab}\,\hat{1}\,,
\end{equation}
(here $\hat1$ is the unit matrix $4 \times 4$)
and a set of spinorial connections,
$\Gamma_a$, associated with a covariant derivative.
Both sets are  matrices $4 \times 4$.

Accepting these notions as fundamental, one may construct theory as follows.
Introduce
dimensionless operators, $\pi_a$,  according to formula,
\begin{equation}\label{pi}
\pi_a\Psi \,=\, -iL(\partial_a - \Gamma_a)\Psi\,.
\end{equation}
Then, one may construct an operator,
\begin{equation}\label{phi_ab}
\phi_{ab}\, =\, \gamma_{[a}\,\gamma_{b]} - \pi_{[a}\,\pi_{b]}\,.
\end{equation}
The last term in (\ref{phi_ab}) is proportional to the curvature tensor,
\begin{eqnarray}
\rho_{ab} & = & 2\pi_{[a}\,\pi_{b]}\\
& = & L^2\left(\partial_a\,\Gamma_b -  \partial_b\,\Gamma_a  - [\Gamma_a,\,
\Gamma_b]\right)\,.
\end{eqnarray}
Next, one may construct a scalar density ($c$--number), using only 
operator $\phi_{ab}$:
\begin{equation}
\phi  =  {1\over 
{4!}}\,e^{abcd}\,e^{efgh}\,{1\over{4}}\,{\rm Tr}\left[\phi_{ae}\,
\phi_{bf}\,\phi_{cg}\,\phi_{dh}\right]\,;
\label{1.2}
\end{equation} 
here the absolute antisymmetric symbol $e^{abcd} =
e^{[abcd]} = \pm 1$. For the action of gravitational field
in absence of other interactions (i.e. of electroweak and strong)
one may take the following expression:
\begin{equation}
S_{g} = K\!\int\! d\Omega [\sqrt{-\phi}\, - \,\sqrt{5}\sqrt{-g}\,]\,.
\label{1.1}
\end{equation}
Here $g = {\rm det}\left[{1\over 4}{\rm Tr}
\left(\gamma_a\,\gamma_b\right)\right]$. It's worthy to mention here, 
that first term in the action (\ref{1.1}) (i.e. $\sqrt{-\phi}$) is 
form-invarant with respect to transformations,
\begin{eqnarray}\label{theta1}
\gamma_a & = & \cosh \theta\,\gamma'_a + \sinh \theta\,\pi'_a\,;\\
\label{theta2}
\pi_a & = & \sinh \theta\,\gamma'_a + \cosh \theta\,\pi'_a\,;
\end{eqnarray}
here $\theta$ is some constant. One may claim this symmetry as fundamental
and demand all terms in the action to be invariant with respect
to (\ref{theta1}) and (\ref{theta2}). 
A theory
of electroweak and gravitational fields, based on this idea,
is constructed in \cite{9608014}; the reader is referred
to this work for details. As it is shown, the characteristic length, 
$L = (20/3)^{1\over2}(k^{1\over2}\hbar)/(ec)$.\footnote{This formula
was obtained by taking into consideration only of electroweak and
gravitational fields; introduction of the strong interaction
should change the numeric
value. The order of magnitude, though, shouldn't change substantially.} 
In absence of other fields and matter, the action
suggested in \cite{9608014} is reduced to (\ref{1.1}).

\section{Vacuum equations for gravitational field}
\setcounter{equation}{0}

Taking variation of (\ref{1.1}) in $\gamma_a$ and $\Gamma_a$,
one obtains the following set of equations:
\begin{eqnarray}\label{1.5}
{1\over{\sqrt{-g}}}\,\partial_a\left[\sqrt{-g}\,\phi^{ba}\,\right]
- \left[\Gamma_a\,, \phi^{ba}\,\right] & = & 0\,;\\
\label{1.6}
\left[\phi^{ab}\,, \gamma_a\,\right] & = & 2\,\gamma^b\,.
\end{eqnarray}
Here $\gamma^b = g^{ba}\gamma_a$ and $g^{ab}$ is a contravariant metric 
tensor ($g_{ab} = {1\over4}\,$Tr($\gamma_a\,\gamma_b$));
the following definitions are used:
\begin{eqnarray}\label{1.7} 
\phi^{dh} & = & {{\sqrt{-\phi}}\over{\sqrt{-5g}}}\varphi^{dh}\,;\\
\label{1.7'} 
\varphi^{dh} & = & 
{1\over{3!\phi}}\,e^{abcd}\,e^{efgh}\,
\phi_{ae}\,\phi_{bf}\,\phi_{cg}\,.
\end{eqnarray}
In \cite{9608014} it is shown that in the limit $L \rightarrow 0$
action (\ref{1.1}) (and hence equations (\ref{1.5}), (\ref{1.6}))
are equivalent to those of Einstein's theory. 

\section{Spherically symmetric static metrics}
\setcounter{equation}{0}

To test equations (\ref{1.5}), (\ref{1.6}), one may apply them to the case
of static, spherically symmetric gravitational field. The difference between
usual approach and the one employed here is the spinorial representation
of basic notions (i.e. of Dirac matrices, $\gamma_a$ and connections, 
$\Gamma_a$). To begin with, one assumes that each component  of $\phi^{ab}$,
defined by (\ref{1.7}),
is proportional to $\gamma^{[a}\,\gamma^{b]}$. (Later it will be
shown that this assumption really takes place.) Then, from (\ref{1.6})
follows,
\begin{eqnarray}
\label{2.1}
\phi^{01} & = & - {m\over{2 + m}}\,\gamma^{[0}\,\gamma^{1]}\,;\\
\label{2.2}
\phi^{23} & = & - 
{m\over{2 + m}}\,\gamma^{[2}\,\gamma^{3]}\,;\\
\label{2.3}
\phi^{0\sigma} & = & - {1\over{2 + m}}\,
\gamma^{[0}\,\gamma^{\sigma]}\,;\;\;\;\sigma = 2, 3;\\
\label{2.4}
\phi^{1\sigma} & = & - {1\over{2 + m}}\,
\gamma^{[1}\,\gamma^{\sigma]}\,.
\end{eqnarray}
Here $m$ is some function.  
{}From (\ref{1.7}), (\ref{2.1}) -- (\ref{2.4}), one obtains, 
\begin{eqnarray}
\label{2.5}
\phi_{01} & = & \alpha_1\,\gamma_{[0}\,\gamma_{1]}\,;\\
\label{2.6}
\phi_{23} & = & \alpha_1\,\gamma_{[2}\,\gamma_{3]}\,;\\
\label{2.7}
\phi_{0\sigma} & = & \alpha_2\,\gamma_{[0}\,\gamma_{\sigma]}\,;\\
\label{2.8}
\phi_{1\sigma} & = & \alpha_2\,\gamma_{[1}\,\gamma_{\sigma]}\,.
\end{eqnarray}
Take for the interval usual expression, namely,
\begin{equation}
ds^2 = e^{2\lambda}dt^2 - e^{2\nu}dr^2 - r^2[d\theta^2 + 
\sin^2\theta\,d\varphi^2]\,.
\label{interval}
\end{equation}
Here $\lambda$ and $\nu$ are functions of $r$ only, and 
$(t, r, \theta, \varphi)$
are spherical coordinates, having traditional interpretation.
Then, for Dirac matrices one can take 
\begin{eqnarray}
\label{2.9}
\gamma_0 & = & e^{\lambda}\,\left( \begin{array}{cc}
1 & 0  \\  
0  &  -1  \end{array} \right) \,;\\
\label{2.10}
\gamma_1 & = & e^{\nu}\,\left( \begin{array}{cc}
0 & \sigma_{1}  \\  
- \sigma_{1}  &  0  \end{array} \right) \,;\\
\label{2.11}
\gamma_2 & = & r\,\left( \begin{array}{cc}
0 & \sigma_{2}  \\  
- \sigma_{2}  &  0  \end{array} \right) \,;\\
\label{2.12}
\gamma_3 & = & r\,\sin\theta\,\left( \begin{array}{cc}
0 & \sigma_{3}  \\  
- \sigma_{3}  &  0  \end{array} \right) \,.
\end{eqnarray}
The following notations are used for linear combinations
of standard Pauli matrices,
\begin{eqnarray}
\label{2.13}
\sigma_1  &  =  &  \left( \begin{array}{cc} 
\cos \theta  &  \sin\theta \exp (- i\varphi )  \\
\sin\theta \exp ( i\varphi ) &  - \cos\theta \end{array} \right) \,; \\
\label{2.14} 
\sigma_2  &  =  &  \left( \begin{array}{cc} 
- \sin\theta  &  \cos\theta \exp (- i\varphi )  \\ 
\cos\theta \exp ( i\varphi )  &  \sin\theta \end{array} \right) \,; \\
\label{2.15} 
\sigma_3  &  =  &  \left( \begin{array}{cc} 
0  &  - i\exp (- i\varphi )  \\ 
i\exp ( i\varphi )  &  0  \end{array} \right) \,.
\end{eqnarray}
Using (\ref{1.5}), (\ref{2.9}) -- (\ref{2.15}), one obtains the following
formulae for connections: 
$\Gamma_0 = a_0 e^{- \lambda - \nu}\gamma_{[0}\,\gamma_{1]}$; 
$\Gamma_1 = 0$; $\Gamma_{\sigma} = {1\over r}a_2 e^{- \nu}
\gamma_{[1}\,\gamma_{\sigma ]}$; $\sigma = 2, 3$. 
Here one uses notations,
\begin{eqnarray}
\label{a_0}
a_0 & = & {1\over2}e^{\lambda - \nu}\left[\lambda' - 
{{m'(1 + m)}\over{2 + m}} - {{m^2 - 1}\over r}\right]\,;\\
\label{a_2}
a_2 & = & {1\over2} - {{e^{- \nu}}\over2}\left(m +
 {{rm'}\over{2 + m}}\right)\,.
\end{eqnarray}
Here prime denotes differentiation over $r$. 
For the curvature tensor components one obtains expressions:
\begin{eqnarray}
\label{rho_01}
\rho_{01} & = & - e^{- \lambda - \nu}{a_0}'\,\gamma_{[0}\,\gamma_{1]}\,;\\
\label{rho_0sigma}
\rho_{0\sigma} & = & {{e^{- \lambda}}\over r}a_0 (2a_2 - 1)\, 
\gamma_{[0}\,\gamma_{\sigma ]}\,;\\
\label{rho_1sigma}
\rho_{1\sigma} & = & {{e^{- \nu}}\over r}{a_2}'\, 
\gamma_{[1}\,\gamma_{\sigma ]}\,;\\
\label{rho_23}
\rho_{23} & = & {2\over{r^2}}a_2 (1 - a_2)\, \gamma_{[2}\,\gamma_{3]}\,.
\end{eqnarray}

{}From (\ref{1.7}), (\ref{2.1}) -- (\ref{2.8}), one obtains algebraic 
relations between $\alpha_1$, $\alpha_2$ and $m$:
\begin{eqnarray}
\label{2.16} 
{{\sqrt{15}m}\over{2 + m}} & = & {{\alpha_1(3\alpha_1^2
 + 2\alpha_2^2)}\over 
{\sqrt{3\alpha_1^4 + 4\alpha_1^2\alpha_2^2 + 8\alpha_2^4}}}\,;\\
\label{2.17} 
{{\sqrt{15}}\over{2 + m}} & = & {{\alpha_2(\alpha_1^2
 + 4\alpha_2^2)}\over
{\sqrt{3\alpha_1^4 + 4\alpha_1^2\alpha_2^2 + 8\alpha_2^4}}}\,.
\end{eqnarray}
Solving eqs. (\ref{2.16}), (\ref{2.17}) with respect to
$\alpha_1$ and $\alpha_2$, one obtains,
\begin{eqnarray}
\label{alpha_1m}
\alpha_1(m) & = & {{\sqrt{15}k}\over{2 + m}}\,
\sqrt{{mk + 2}\over{k^2 + 4}}\,;\\
\label{alpha_2m}
\alpha_2(m) & = & {{\sqrt{15}}\over{2 + m}}\,
\sqrt{{mk + 2}\over{k^2 + 4}}\,;
\end{eqnarray}
here $k = {\alpha_1}/{\alpha_2}$ is a real root
of equation,
\begin{equation}\label{k_eq}
k^3 - {m\over3}k^2 + {2\over3}k - {{4m}\over3}\, =\, 0\,;
\end{equation}
The root can be written explicitly as 
\begin{equation}
\label{k}
k = {1\over9}m + {1\over3}A(m) + {1\over3}B(m)\,,
\end{equation}
where
\begin{eqnarray}
\label{A}
A(m) & = & \sqrt[3]{17m + {1\over{27}}m^3 + \sqrt{8 + {{863}\over3}m^2 +
{4\over3}m^4}}\,;\\
\label{B}
B(m) & = & \sqrt[3]{17m + {1\over{27}}m^3 - \sqrt{8 + {{863}\over3}m^2 +
{4\over3}m^4}}\,.
\end{eqnarray}

{}From (\ref{phi_ab}), (\ref{2.5}) -- (\ref{2.8}), (\ref{rho_01}) --
(\ref{rho_23}), one obtains after some manipulation with formulae,
expressions for $\lambda$ and $\nu$ as functions of $r$, $\alpha_1$,
$\alpha_2$, $m$:
\begin{eqnarray}
\label{e_nu}
e^{\nu} & = & {{L^2}\over{4r(\alpha_2 - 1)}}\,\left[\sqrt{1 -
{{4r^2}\over{L^2}}(1 - \alpha_1)}\,\right]'\,;\\
\label{lambda'}
\lambda' & = & \left[{1\over2}\ln\left(1 - 
{{4r^2}\over{L^2}}(1 - \alpha_1)
\right) + m - \ln (2 + m)\right]' + {{m^2 - 1}\over {r}}\,.
\end{eqnarray}
Further, there are two differential equations, imposed on functions
in case:
\begin{eqnarray}
\label{beta}
{{r{\alpha_1}'}\over{2(\alpha_2 - 1)}} - {{rm'}\over{2 + m}} 
 - m + {{\alpha_1 - 1}\over{\alpha_2 - 1}} & = & 0\,;\\
\label{aux}
{{r{\alpha_2}'}\over{(\alpha_2 - 1)}} + {{rm'}\over{2 + m}} +
\left({{rm'}\over{2 + m}} + m\right)\!\left(m - 
{{\alpha_1 - 1}\over{\alpha_2 - 1}}\right) & = & 0\,.
\end{eqnarray}
It is easy to show that compatibility condition for (\ref{beta})
and (\ref{aux}) can be written as
\begin{equation}\label{compatible}
\dot{\alpha}_2 + {m\over2}\dot{\alpha}_1
+ {{\alpha_2 - \alpha_1}\over{2 + m}}\, =\, 0\,,
\end{equation}
where dot denotes differentiation
over $m$. This condition actually holds
provided that (\ref{alpha_1m}) -- (\ref{k_eq}) take place; thus one
may consider only eq. (\ref{beta}), disregarding  (\ref{aux}).

Solving (\ref{beta}), one obtains,
\begin{equation}
\label{r}
r\,=\,r_{\star}\,\sqrt[3]{|2 + m|}\, 
|m(\alpha_2 - 1) - (\alpha_1 - 1)|^{-{1\over{2 + m^2}}}F(m)\,,
\end{equation}
where $r_{\star}$ is a constant of integration, and
\begin{eqnarray}
\label{F_m}
F(m) & = & {1\over
{\sqrt[6]{2 + m^2}}}\,\exp\left(
-{1\over{3\sqrt{2}}}\arctan {m\over{\sqrt{2}}}\right)
\exp\left[-\int_{1}^m\!dm'M(m')\,\right]\,;\\
\label{M_m}
M(m)  & = & {{2m}\over{(2 + m^2)^2}}\ln
|m(\alpha_2 - 1) - (\alpha_1 - 1)|\,.
\end{eqnarray}
Inverting (\ref{r}), one obtains function $m(r)$.
Function $F(m)$ in (\ref{F_m}) `behaves properly', i.e. it doesn't
have zeroes and poles in finite range of $m$. In the case $m'\,\neq\,0$,
i.e. when formula (\ref{r}) takes place (case $m =$ constant will be
considered separately), one may obtain the following formulae for
metrics coefficients, using (\ref{e_nu}) -- (\ref{beta}),
(\ref{r}),
\begin{eqnarray}
\label{g00_reduced}
g_{00} & = & {{r_{\star}^6}\over{r^6}}\left[1 + {{4r^2}\over{L^2}}(\alpha_1 - 1)
\right]{{(2 + m)^2}\over{[m(\alpha_2 - 1) - (\alpha_1 - 1)]^2}}\,;\\
\label{g11_reduced}
g_{11} & = & - \left[1 + {{4r^2}\over{L^2}}(\alpha_1 - 1)\right]^{-1}
{{[m(2 + m)\dot{\alpha}_1 - 2(\alpha_1 - 1)]^2}\over
{[(2 + m)\dot{\alpha}_1 - 2(\alpha_2 - 1)]^2}}\,.
\end{eqnarray}
Equations (\ref{g00_reduced}), (\ref{g11_reduced}), (\ref{r}) (together with 
respective definitions) give formal solution to the problem. 
One should bring to attention,
though, the fact that function $r(m)$ is not monotonic; so one has to
cut interval $(-\infty,\,+\infty)\,\ni\,m$ into domains of monotony 
of the function $r(m)$. Each such domain would correspond to some 
solution to the problem. One may distinguish the following intervals:
(i) $(-\infty,\,-2)$; (ii) $(-2,\,m_0)$; (iii) $(m_0,\,m_{min})$;
(iv) $(m_{min},\,1)$; (v) $(1,\,+\infty)$. Here $m_0\,\approx\,-1.60808367$;
$m_{min}\,\approx\,-1.365056$. 
Function $r(m)$ behaves monotonically in each interval.
Consider each interval separately. 
 
{\underline{Case $m \in (1,\,+\infty)$: 
The Schwarzschild solution, negative mass}}

On interval function $r(m)$ decreases from $+\infty$ to a constant value.
At the left border of interval $r\,\rightarrow\,\infty$,
metrics corresponds to the Schwarzschild solution with $r_g < 0$.

{\underline{Case $m \in (m_{min},\,1)$: 
The Schwarzschild solution, positive mass}}

Interval corresponds to an exterior solution. Function $r(m)$ 
increases from minimal value $r\,=$ constant to $+\infty$.
 At the right border of the interval $(m\,\rightarrow\,1)$ one may expand
all functions in case into series in small parameter $(1 - m)$; inverting
the expansion for $r(m)$, one may find $m(r)$. Omitting details of
computation,
one may present the following formulae for metrics' expansion in parameter
$L$  (up to terms of order $L^4$):
\begin{eqnarray}
\label{g_00}
g_{00} & = & \left(1 + {{3r_*^6}\over {r^6}}\right)
\left[1 - {{r_g}\over r}\left(1 - {{r_*^6}\over{r^6}}\right)\right]\,;\\
\label{g_11}
g_{11} & = & - \, \left(1 - {{9r_*^6}\over {r^6}}\right)
\left[1 - {{r_g}\over r}\left(1 - {{r_*^6}\over{r^6}}\right)\right]^{-1}\,.
\end{eqnarray}
Here the following notation is used:
\begin{equation}
\label{r_*}
r_*^6 \, = \, {{r_g^2 L^4}\over {80}}\,.
\end{equation}
Here $r_g$ is another constant, corresponding to the
Schwarzschild radius of Einstein's theory. Constant $r_{\star}$ 
is connected to $r_g$ by relation,
\begin{equation}
\label{r_star}
r_{\star}^3\,=\,p_1r_gL^2\,,
\end{equation}
where
\begin{equation}\label{p1}
p_1\,=\,{{\sqrt{3}}\over8}\exp\left({1\over{\sqrt{2}}}\arctan{1\over
{\sqrt{2}}}\right)\,\approx\,0.3345645\,.
\end{equation}
Due to (\ref{r_*}) corrections of order $L^2$ are
absent in metrics' decomposition. The domain of validity for (\ref{g_00}),
(\ref{g_11}) is $r\, \gg\, r_*$.
At the left border, $m\,=\,m_{min}$, $r(m)$ achieves minimum; 
one may expand, again, $r(m)$ into
series near minimal value; thus,
\begin{equation}\label{r_to_r_star}
{r\over{r_{\star}}}\,=\,A + B(m - m_{min})^2 + \cdots
\end{equation}
Numeric computations give $A\,\approx\,0.48024254$, 
$B\,\approx\,0.60170272$. Inverting 
(\ref{r_to_r_star}) and substituting into (\ref{g00_reduced}),
(\ref{g11_reduced}), one obtains the following asymptotics:
\begin{eqnarray}\label{g00_approx1}
g_{00} & = & 15.144 \times f(m_{min}) + O\left({r\over{r_{min}}} - 
1\right)\,;\\
\label{g11_approx1}
g_{11} & = & - 0.495 \times [f(m_{min})]^{-1}\left({r\over{r_{min}}}
- 1\right)^{-1} + O(1)\,.
\end{eqnarray}
Here one uses the notation,
\begin{equation}\label{f1}
f(m)\,=\,1 + {{4r^2}\over{L^2}}(\alpha_1 - 1)\,.
\end{equation} 
Value $r_{min}\,\equiv\,r(m_{min})\,=\,A r_{\star}$ (c.f. 
(\ref{r_to_r_star})). Comparing this
expression with (\ref{r_star}), one obtains,
\begin{equation}\label{f}
f(m_{min})\, =\, 1 - q_0\left({{r_g}\over{L}}\right)^{2\over3}\,;
\end{equation}
here $q_0\,\approx\,3.0767137$. Using numerical computations, one may
show that $f(m)$ monotonically increases on interval $m\in (m_{min},\,1)$.
One also has $f(1) = 1$.
This implies that formation of a horizon (corresponding to $f(m_g) = 0$ 
for
some $m_g\in (m_{min},\,1)$) depends on sign of $f(m_{min})$. Namely,  
horizon forms if and only if $f(m_{min}) < 0$. 
According to (\ref{f}), masses
$r_g < 0.185 L$ don't form the horizon in the domain of validity, i.e.
in the region $r > r_{min}
\approx 0.333(r_gL^2)^{1\over3}$. Note, that in this case $r_{min}$
doesn't exceed $0.2L$.
The coordinate system $(t,\,r,\,\theta,\,\varphi)$ cease to be valid in
the region $r < r_{min}$. Discussion of some details of the solution
will be postponed until the conclusion.

{\underline{Case $m \in (m_0,\,m_{min})$: Interior solution I}}

Function $r(m)$ decreases from $r\,=\,+\infty$ to $r\,=$ constant, 
achieving minimum.
On the right border of interval formulae (\ref{g00_approx1})
and  (\ref{g11_approx1}) are still valid. One has $\dot{f}(m) > 0$ 
(respectively, $f'(m) < 0$) on the interval. Besides, $f(m_0) = -\infty$.
This means, that if $f(m_{min}) > 0$, then solution with correct signature
($g_{00} > 0$, $g_{11} < 0$) exists only for $r_{min} < r < r(m_{u})$, where
$m_{u}$ is determined by equation $f(m_{u}) = 0$. Note, that since the 
Schwarzschild solution is not linked with the one in consideration,
one cannot use
formula (\ref{r_star}); in this case $r_{\star}$ is independent constant.

{\underline{Case $m \in (-2,\,m_0)$: Interior solution II}}

Function $r(m)$ increases from $r\,=\,0$ to $r\,=\,+\infty$.
At the left border of interval,
asymptotics for the metric coefficients is the following:
\begin{equation}\label{6.1}
g_{00} \, \propto \, {{r^2}\over{r_*^2}}\,;\;\;\;
g_{11} \, \propto \, - {{r^4}\over{r_*^4}}\,.
\end{equation}
Here $r_*$ is a constant of integration. On the interval $\dot{f}(m) < 0$;
as it was mentioned above, $f(m_0) = -\infty$. One has a freedom to choose
the constant of integration ($r_{\star}$ in (\ref{r})), so that $f(-2) > 0$.
Thus, the solution exists for $0 < r < r_{m}$. 

{\underline{Case $m \in (-\infty,\,-2)$: Interior solution III}}

Function $r(m)$ decreases from $r\, =$ constant to $r\,=\,0$.
At the right border of interval 
asymptotics (\ref{6.1}) is
still valid. $f(m) > 0$ everywhere on interval, so the metrics has
correct signature for $0 < r < r_{max}$, where $r_{max} = r(-\infty)$.  

\section{Another interior solution}
\setcounter{equation}{0}

In obtaining the solutions above it was assumed 
that $m$ is a functions of $r$. 
One may also search for solutions with $m$
to be constant. Assuming so, one obtains from (\ref{beta}) 
an equation for $m$:
\begin{equation}
\label{m_0}
{{\alpha_1(m) - 1}\over{\alpha_2(m) - 1}}\, = \, m\,. 
\end{equation} 
The numerical computation of (\ref{m_0}) gives the root,
$m_0 \approx - 1.60808367$; respectively, $\alpha_1(m_0)
\approx - 10.8671667$.
The solution for metrics, due to (\ref{e_nu}) and (\ref{lambda'}), is:
\begin{equation}
ds^2 \, = \, \left({{r_0^2}\over{r^2}} - 1\right)
\left({r\over{r_0}}\right)^{2m_0^2}\!c^2 dt^2 -
{m_0^2}\left(1 - {{r^2}\over{r_0^2}}\right)^{-1}\!\!\!dr^2  
- r^2\left(d\theta^2 +
\sin^2\theta\,d\varphi^2\right)\,.
\label{bubble}
\end{equation}
Here
\begin{equation}
r_0 \, = \, {L\over{2\sqrt{|\alpha_1(m_0)| + 1}}}\,.
\end{equation}
Solution (\ref{bubble}) makes sense
for $r\,\in\,[0,\,r_0)$.

\section{Conclusion}
\setcounter{equation}{0}

Among static, spherically symmetric solutions for metrics, 
predicted by the Born--Infeld theory, the most 
interesting is the exterior one, having asymptotic of the
Schwarzschild solution on $r \rightarrow \infty$. This solution
exhibits at small radial distances behavior, dramatically 
different from those of the traditional solution for a black hole. 

Opposite to a black hole solution with a spacetime singularity, 
(a timelike curve $r = 0$), one deals now with a hypersurface, 
$r = r_{min} \approx 0.333 \times (r_g L^2)^{1\over3}$,
which presents a boundary to the solution.
For $r < r_{min}$ the solution (\ref{g00_reduced}), (\ref{g11_reduced})
cease to be valid, since in this region determinant of the metric
tensor, calculated with (\ref{g00_reduced}), (\ref{g11_reduced}), $g > 0$. 
This circumstance doesn't exclude the region from the
physical description. It means, instead, that one has to solve field 
equations
(\ref{1.5}), (\ref{1.6}) (modified to include the matter), for
solution corresponding to the interior of the mass, and sew both
interior and exterior solutions on the boundary. One should emphasize, though,
a remarkable fact, that the exterior solution ``leaves a vacancy'' (i.e.
a spacial volume), $V \propto r_{min}^3$, for the mass. 
This actually means that matter cannot be squeezed (even by forces
other than gravity) beyond the radius $r_{min}$, otherwise the field
equations wouldn't be consistent.\footnote{One may guess, that the minimal
radius, $r_{min}$, will change after `switching on' other interactions;
though, one expects that general structure of the solution will remain the
same.}
Using dimensional
analysis, one obtains the average density of the mass, 
$\rho \propto M/r_{min}^3$; the coefficient of proportionality
depends on geometry of space inside the volume. This density doesn't 
depend on the value of the mass, $M$, and, in fact, is of order of 
magnitude
of the Planck's density. Thus, Born--Infeld gravity is more `benign'
than that of Einstein: it doesn't squeeze matter more than to the
Planck's density (approximately).
Unlike a black hole
singularity, 
the boundary $r = r_{min}$ doesn't have
physical infinities on it. Really, 
the curvature invariants, $R^{abcd}R_{abcd}$, and
$R_{\;\;\;ab\,}^{\,cd}R_{\;\;\;eh\,}^{\,ab}R_{\;\;\;cd}^{\,eh}$
are finite on the boundary.
Here $R_{\,bcd}^{\,a}$ is the Riemann tensor, 
calculated with metrics 
(\ref{g00_approx1}) and (\ref{g11_approx1}). In short, Born--Infeld theory
replaces a black hole's point-like spacial singularity with infinite 
density of mass, by
a `ball of matter' with finite density of order of magnitude
of the Planck's density. 
The same solution, (\ref{g00_reduced}), (\ref{g11_reduced}) may serve
as an exterior part in the case when the mass has lesser than `ultimate'
density; one should, then, sew solutions (both exterior and interior for
the case) together at some $r > r_{min}$.

Another notable difference, is that surface of a horizon, which is
defined in Einstein's theory by equation $r = r_g$, in Born--Infeld
theory should be defined by equation $r = r_h$, where $r_h = r(m_h)$,
and $m_h$ is the root of equation $f(m_h) = 0$ (c.f. (\ref{f1})). 
For example, for $r_g \gg r_*$,
one obtains $r_h \approx r_g - r_*^6/r_g^5$, where $r_*$ is given
by (\ref{r_*}).

For masses $M \gg 10^{-5}$ cm, spacetime geometry 
is similar to that described by the Schwarzschild solution in a sense that
the horizon exists at $r \approx r_g \gg 
r_{min}$. As the mass, $M$, decreases, $r_h$ decreases faster than
$r_{min} \propto \sqrt[3]{r_g}$, so that both surfaces (the horizon and
the boundary) fuse at $r \approx 0.185 L$. Further decrease of
the mass leads to disappearance of the horizon, 
so that the boundary becomes ``naked''.  

It makes sense to consider an interaction of two microscopic, 
`ultimately squeezed' masses, gravitating with each other according to
the classical potential (c.f. (\ref{g_00})).
Take masses, $M \approx
10^{-20}$ g. Then, from (\ref{g_00}) follows, that at sufficiently
large distances, the  potential, $V(r) \sim 
{{3r_*^6}\over{r^6}} - {{r_g}\over{r}}$. For this case, $r_g \sim
10^{-15} L$, and $r_* \sim (r_g/L)^{1\over3}L \sim 10^{-5}L$. 
(From (\ref{r_*}) follows, $r_* \approx 0.482 (r_gL^2)^{1\over3}$,
so that $r_* > r_{min}$.)
The
distance, $r_{eq}$, at which potential energy $V(r)$ has minimum,
$r_{eq} \sim
(r_g/L)^{1\over5}L \sim 10^{-3}L \sim 10^{-36}$ cm. 
Note, that $r_*/r_{eq} \sim 10^{-2}$, so using of formula (\ref{g_00}) 
is justified.        
Thus, 
for $r > r_{eq}$,
the mass attracts, and on distances
$r < r_{eq}$, it repulses. This repulsion may prevent fusion of
two (or more) `particles' with sufficiently small masses, 
or at least make such fusion less probable. On the other hand, for
masses with $r_g > 0.185L$, the horizon exists. 
One may guess, that such a mass, undergoing 
implosion and passing beyond
the horizon, will be squeezed up to the boundary, i.e. will form
a core with average density of order of magnitude of the Planck's density.

The interior solution, (\ref{bubble}), is unique in a sense that it
has fixed, well defined dimensions. This solution corresponds to a
closed space with finite volume, $V = 2|m_0|\pi^2r_0^3$. This
microscopic universe might be considered as a candidate for 
a ``seed'', which could
inflate under certain circumstances (Big Bang) into a Universe,
similar to ours.  

\section{Acknowledgements}

I'm grateful to my friend, Prof. B. S. Tsirelson, who has spent some time 
programming the ``Maple'' for numeric computations; 
without his help it would take much longer to complete this work.

\end{document}